\documentclass[aps,prb,10pt,twocolumn,letterpaper,amsmath,amssymb]{revtex4-2}

\newcommand{\beginsupplement}{%
        \setcounter{table}{0}
        \renewcommand{\thetable}{S\arabic{table}}%
        \setcounter{figure}{0}
        \renewcommand{\thefigure}{S\arabic{figure}}%
        \setcounter{equation}{0}
        \renewcommand{\theequation}{S\arabic{equation}}%
        \setcounter{section}{0}
     } %\beginsupplement

\usepackage[utf8]{inputenc}
\usepackage{graphicx}
\usepackage{subfigure}
\usepackage[usenames,dvipsnames]{xcolor}
\usepackage{epstopdf}
\usepackage{amsmath,amsthm,amssymb}
\usepackage{tcolorbox}
\usepackage{latexsym}
\usepackage{bm}
\usepackage{dcolumn}
\usepackage{braket}
\usepackage[normalem]{ulem}
\usepackage{times}
\usepackage{hyperref}
\usepackage{enumitem}
\newcommand{\ignore}[1]{}

\newcommand{\boxedpp}{\boxed{{}_+^+}}
\newcommand{\boxedpm}{\boxed{{}_+^-}}

\newcommand{\JF}{J}
\newcommand{\JAF}{J_\mathrm{AF}}
\newcommand{\newa}{\widetilde{\mu_a}}
\newcommand{\newb}{\widetilde{\mu_b}}
\newcommand{\newab}{\widetilde{\JAF}}
\newcommand{\newaa}{\widetilde{\JF}}
\renewcommand{\eqref}[1]{Eq.~(\ref{#1})}
\DeclareMathOperator{\csch}{csch}

\newcommand{\be}{\begin{equation}}
\newcommand{\ee}{\end{equation}}
\newcommand{\bea}{\begin{eqnarray}}
\newcommand{\eea}{\end{eqnarray}}

\newcommand{\mytitle}{Metamagnetic Transition in Low-Dimensional Site-Decorated Quantum Heisenberg Ferrimagnets}

\begin{document}
\title{\mytitle}
\author{Weiguo Yin}
%\email{wyin@bnl.gov}
\author{A. M. Tsvelik}
%\email{atsvelik@bnl.gov}
\affiliation{Condensed Matter Physics and Materials Science Division,
Brookhaven National Laboratory, Upton, New York 11973, USA}
\date{\today}

\begin{abstract}
The prohibition of finite-temperature phase transition in one-dimensional (1D) Ising models and 1D/2D quantum Heisenberg models with
short-range interactions fundamentally constrains the application potentials of low-dimensional magnetic materials. Recently, ultranarrow phase crossover (UNPC), which can approach a transition at a desirable finite temperature $T_0$ arbitrarily closely, was discovered in 1D decorated Ising chains and ladders. Here we present a theoretical study of similarly decorated, yet much more challenging, quantum Heisenberg ferrimagnets in a magnetic field, which features ferromagnetic backbone exchange $J$, antiferromagnetic site-decoration coupling $\JAF$, and different magnetic moments for the backbone and decorating spins $\mu_aS_a<\mu_bS_b$. We exactly solved the model in the large $J$ limit---as a central-macrospin model---and found two finite-temperature second-order transitions; just above $T_{c2}$ a ``half-ice, half-fire'' regime appears. Finite-$J$ weak-field results follow from an effective-field mapping, suggesting the emergence of UNPC at finite $T_0$ in 2D square lattices thanks to its exponentially strong initial magnetic susceptibility $\chi_0\propto e^{4\pi S_a^2 J/T_0}$, though less likely in 1D chains where $\chi_0\propto J/T_0$. These results may shed light on new technological applications of low-dimensional quantum spin systems and attract experimental and computational tests.
\end{abstract}

\maketitle

%\section{Introduction}
\emph{Introduction.---}Finding %novel 
states with new functionalities and how they switch to one another %and phase transitions 
is a central problem in materials science~\cite{Kivelson_24_book_statistical,Mattis_book_08_SMMS}. The states that may play important roles in quantum computing, spintronics, unconventional superconductivity, and magnetic refrigeration exist %abound 
in frustrated magnets %which give rise to such exotic magnetic states 
as spin liquid, spin ice, spin glass, spin supersolid, and skyrmion %that may play important roles in quantum computing, spintronics, and unconventional superconductivity
~\cite{Balents_nature_frustration,Miyashita_10_review_frustration,NNano_13_review_skyrmion,PRXEnergy_24_magnetocaloric}. Frustrated magnets also provide a rigorous test ground for machine learning and AI reasoning ~\cite{ML_Hopfield_82,Fan_NC_23_ML-2D-Ising-glass_ML,Yin_Potts_J1-J2_1D}. 
%The essence of frustration is usually assumed to be the ground-state degeneracy that emerges as a consequence of competing exchange interactions among the spins. 
Recently, the prohibition of finite-temperature phase transition in one-dimensional (1D) Ising and Potts models with short-range interactions~\cite{Ising1925,Potts_1952} was circumvented by the emergence of ultranarrow phase crossover (UNPC) when certain frustration was introduced spontaneously~\cite{Yin_MPT} or by an external field via a hidden ``half ice, half fire'' state~\cite{Yin_MPT_chain,Yin_Ising_III_PRL,Yin_g,Yin_site_arXiv,Yin_Potts_UNPC}. The UNPC approaches a genuine transition arbitrarily closely at desirable finite temperature, a step further than the pseudotransition that approaches a zero-temperature transition~\cite{016_Strecka_book_chapter,017_Krokhmalskii_PA_21_3-previous-chains_effective_model}. These findings have opened not only new possibilities of 1D systems for technological applications but also raised a critical question: Can UNPC exist in similarly frustrated 1D and 2D quantum Heisenberg models with short-range interactions, where finite-temperature phase transition is prohibited by the Mermin-Wagner theorem~\cite{Mermin_PRL_theorem}? 

The challenge of this question is clear: Unlike the 1D Ising and Potts models (with one-component spins), even the simplest quantum or classical Heisenberg models (with three-component spins) in the presence of a finite magnetic field~\cite{Fisher_64_1D_Heisenberg_classical,Blume_PRB_75_1D_Classical_Heisenberg_field,Takahashi_PTPS_71_1D_quantum_Heisenberg_h} or frustration~\cite{Harada_ZPB_88_1D_J1-J2_Classical_Heisenberg} have not been solved explicitly---in the sense that a closed-form exact solution to the partition function at finite temperature is found.

\begin{figure}[t]
\includegraphics[width=0.6\columnwidth,clip=true,angle=0]{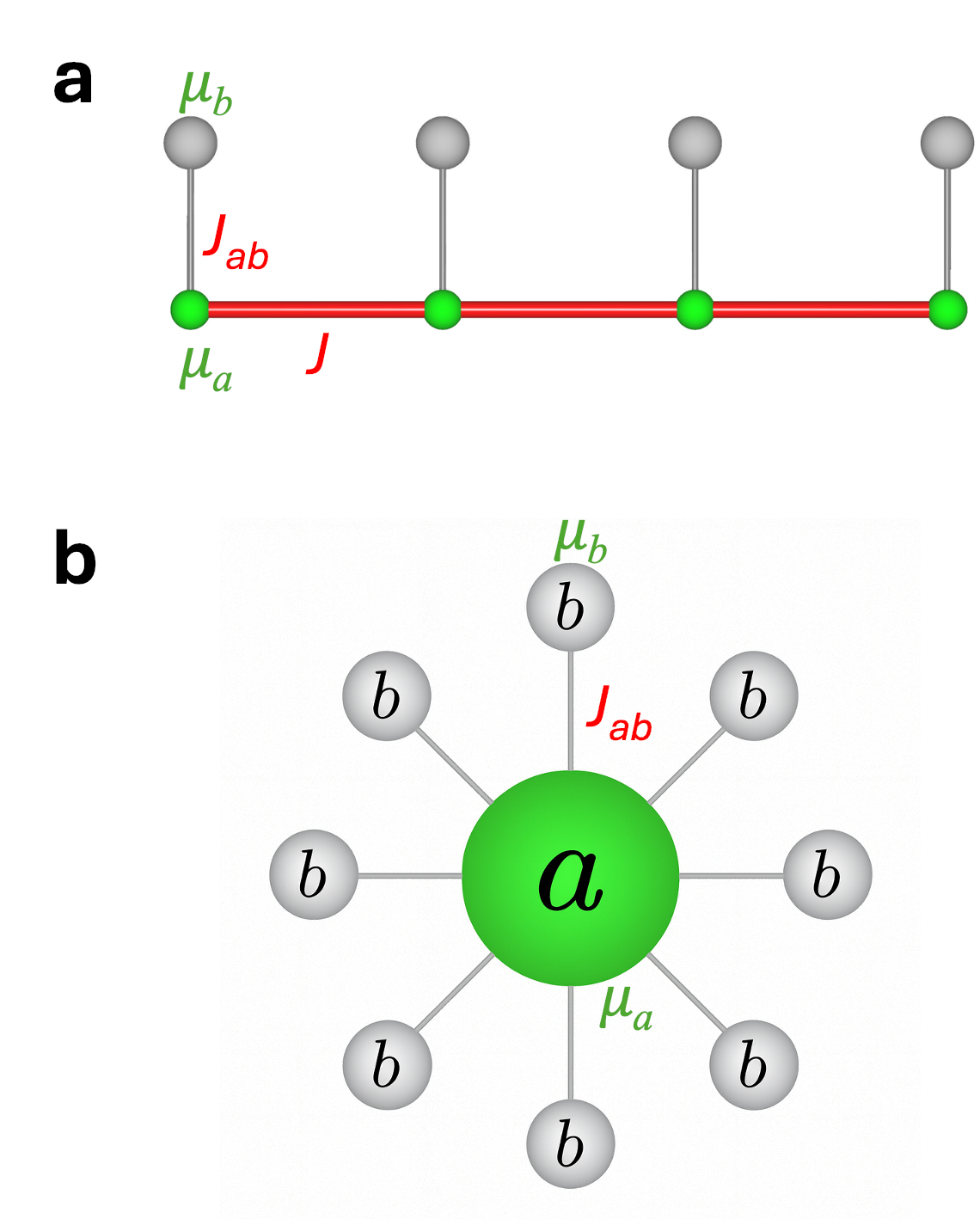}      
\caption{(a) The site-decorated model, where the $a$- and $b$-spins are the backbone and decorating spins, respectively. (b) The central-macrospin model, where the $a$-spins collapse to form one macroscopic spin in the  large $J$ limit of the site-decorated model and the $b$-spins provide a bath environment.}
\label{Fig:model}
\end{figure}

In this paper, with AI-assisted reasoning, we address this question by studying a frustrated Heisenberg model, whose 1D Ising counterpart is known as a minimal model for UNPC driven by the half-ice, half-fire state~\cite{Yin_site_arXiv}. 
%\section{Model Hamiltonian}
%\emph{Model Hamiltonian.---}
The model is a site-decorated ferrimagnet in a magnetic field [Fig.~\ref{Fig:model}(a)]:
\begin{equation}
H=H_a+H_b, \label{eq:model}
\end{equation}
with
\begin{subequations}
\begin{eqnarray}
H_a&=&-\JF\sum_{\langle ij\rangle}\mathbf{S}_{i,a}\cdot\mathbf{S}_{j,a}-h\mu_a\sum_i S_{i,a}^z, \label{eq:Ha}\\
H_b&=&\;\;\JAF\sum_{i}\mathbf{S}_{i,a}\cdot\mathbf{S}_{i,b}-h\mu_b\sum_i S_{i,b}^z, \label{eq:Hb}
\end{eqnarray}
\end{subequations}
where $H_a$ describes the backbone spins $\mathbf{S}_{i,a}$ located at the lattice site $\mathbf{R}_i$ (green balls, referred to as $a$-spins) with the ferromagnetic interaction $\JF>0$ between nearest neighboring $a$-spins (red bonds), and $H_b$ describes the decorating spins $\mathbf{S}_{i,b}$ (gray balls, referred to as $b$-spins) coupled to $\mathbf{S}_{i,a}$ with the antiferromagnetic interaction $\JAF>0$ (gray bonds). $|\mathbf{S}_{i,a}|=\sqrt{S_a(S_a+1)}$ and $|\mathbf{S}_{i,b}|=\sqrt{S_b(S_b+1)}$. $h>0$ is the magnetic field. $\mu_a$ and $\mu_b$ are the Land\'{e} $g$-factors of the $a$- and $b$-spins, respectively. The relationship of $\mu_b S_b > \mu_aS_a > 0$ is used to represent ferrimagnetism. $N$ is the total number of unit cells.
%with $\mathbf{S}_{N+1}\equiv\mathbf{S}_{1}$, $\mathbf{T}_{N+1}\equiv\mathbf{T}_{1}$, viz., 
We use the periodic boundary condition and the natural units of the Bohr magneton $\mu_\mathrm{B}=1$ and the Boltzmann constant $k_\mathrm{B}=1$; thus, $\beta=1/T$.
The 1D Ising or Potts counterpart of this model can be mapped exactly onto a zero-field bond-decorated $J_1$-$J_2$ Ising or Potts chain, unambiguously elucidating the collective nature of the UNPC in a magnetic field~\cite{Yin_site_arXiv,Yin_Potts_UNPC}.

Our goal is to find a sharp metamagnetic transition or UNPC at which the backbone spins flip from pointing down to pointing up---as temperature increases. The phase crossover width in the site-decorated Ising or Potts model is exponentially narrowed as $J$ increases, reaching a transition for $J\to\infty$~\cite{Yin_site_arXiv,Yin_Potts_UNPC}. Thus, we are motivated to study the large $J$ limit of the quantum Heisenberg case, which  becomes \emph{a central-macrospin model} and can be solved exactly for any spatial dimensions of the original model. We find that the half-ice, half-fire state persists in this model. For finite $J$, since the effective field acting on the backbone spins flips its sign, i.e., $h_\mathrm{eff}=0$ at the metamagnetic transition, we can use the known zero-field---which means $h_\mathrm{eff}=0$ here while $h\ne0$---magnetic susceptibility $\chi_0$ of 1D and 2D quantum and classical Heisenberg models, \eqref{eq:Ha}~\cite{Fisher_64_1D_Heisenberg_classical,Takahashi_PRB_87_2D-Classical-Heisenberg,Takahashi_PRL_87_1D2D-quantum-Heisenberg,Takahashi_PTPS_86_1D2D_quantum_Heisenberg}, to evaluate the crossover width $2\delta T$. We predict that, as $J$ increases in a weak field, $2\delta T$ is narrowed in a power law in the 1D chain but exponentially in the 2D square lattice; therefore, UNPC would exist in  decorated square-lattice classical and quantum Heisenberg ferrimagnets in a weak external magnetic field.

%\subsection{$T=0$}
\emph{Ground-state phase diagram.---}In the parameter space of interest where $J$ is considerably stronger than $\JAF$, the zero-temperature ($T=0$) phase diagram of the original quantum Heisenberg model, \eqref{eq:model}, in a finite magnetic field can be obtained in linear spin-wave theory by minimizing the total energy with respect to the polar angles $\theta_a$ and $\theta_b$ of the quantization axes of both $a$- and $b$-spins (or simply the polar angles of $a$- and $b$-spin vectors in the classical-spin framework). Let $\theta_a^*$ and $\theta_b^*$ denote the results of this minimization. 
One finds 
\begin{eqnarray}
&\cos\theta_a^*&=\left\{
\begin{array}{cl}
     -1 & \mathrm{for}\;\; 0<h\le h_{c1},\\
     \frac{h^2-h_{c1}h_{c2}}{h(h_{c2}-h_{c1})} & \mathrm{for}\;\; h_{c1}\le h \le h_{c2},\\
      1 & \mathrm{for}\;\; h\ge h_{c2},\\
\end{array}\right.\label{eq:ma0} 
\end{eqnarray}
where the two critical fields are given by
\begin{eqnarray}
    h_{c1}&=&\JAF S_aS_b\left(\frac{1}{\mu_aS_a}-\frac{1}{\mu_bS_b}\right),\nonumber\\
    h_{c2}&=&\JAF S_aS_b\left(\frac{1}{\mu_aS_a}+\frac{1}{\mu_bS_b}\right). \label{eq:hc12}
\end{eqnarray}
$\theta_b^*$ is given by the torque-balance equation 
\begin{equation}
\mu_bS_b \sin \theta_b^*= \mu_aS_a \sin \theta_a^* \label{eq:mb0}
\end{equation}
with $\cos\theta_b^*=1$ for $\cos\theta_a^*=\pm 1$. The $a$- and $b$-spins are antiferromagnetically and ferromagnetically aligned along the $z$ axis in weak fields $h\le h_{c1}$ and strong fields $h\ge h_{c2}$, respectively. For $h_{c1}\le h \le h_{c2}$, they are canted from each other by the angle $\theta_a^*+\theta_b^*$---while $\cos\theta_b^*$ stays close to $+1$, $\cos\theta_a^*$ continuously increases from $-1$ to $+1$ with $\cos\theta_a^*=0$ occurring at $h=\sqrt{h_{c1}h_{c2}}$ (see Fig.~\ref{Fig:angle}(a) in Supplemental Material~\cite{SI}). The canted phase is invariant with respect to rotation around the $z$ axis; therefore, the spin waves are gapped for $h< h_{c1}$ and $h> h_{c2}$ but gapless for $h_{c1}\le h \le h_{c2}$ [Fig.~\ref{Fig:angle}(b)]. The $T=0$ quantum phase transitions at $h_{c1}$ and $h_{c2}$ are second-order. None of the three phases %in the ground-state phase diagram 
has a macroscopic degeneracy, rendering this system seemingly normal within the traditional treatment of frustrated magnets~\cite{Balents_nature_frustration,Miyashita_10_review_frustration,016_Strecka_book_chapter,017_Krokhmalskii_PA_21_3-previous-chains_effective_model}. However, like its Ising and Potts counterparts, it is a novel unconventional frustrated magnet---lacking the conventional geometric frustration---where the main physics is driven by an excited state with macroscopic degeneracy, which is hidden in the ground-state phase diagram~\cite{Yin_Ising_III_PRL,Yin_site_arXiv,Yin_Potts_UNPC}, as demonstrated below.

%\section{Coronaspin model for $J\to\infty$}
\emph{Finite $T$, infinite $J$.---}In the $J\to\infty$ limit, the $a$-spins form one macroscopic spin $\mathbb{S}_a$ with the magnitude $NS_a$. The site-decorated lattice model in any spatial dimension is reduced to a ``spin star'' structure, which consists of a central macrospin $\mathbb{S}_a$ coupled to a bath of $N$ $b$-spins [Fig.~\ref{Fig:model}(b)]: 
\begin{eqnarray}
H_\mathrm{CMM}=&-&h\mu_a \mathbb{S}_{a}^z \nonumber \\
&+&\JAF\frac{1}{N}\mathbb{S}_{a}\cdot\sum_{i=1}^N\mathbf{S}_{i,b}-h\mu_b\sum_{i=1}^N S_{i,b}^z, \label{corona}
\end{eqnarray}
The model looks like the central-spin model or the spin-star network for quantum information~\cite{ProkofevStamp2000_spin-bath,Villazon_PRR_20_central-spin,BreuerBurgarthPetruccione2004} %,Fauseweh2017ClassicalCSM} 
which focuses on the quantum dynamics of the central spin---as a quantum dot---and how to control the decoherence caused by the spin bath environment as an essential step in constructing qubits for quantum computers. By sharp contrast, the present central macrospin becomes a classical spin in the thermodynamical limit $N\to\infty$. This central-\emph{macrospin} model (CMM) can be solved exactly, particularly in closed form for $S_b=1/2$. Note that the CMM for $J\to\infty$ applies to site, bond, and other decorated Heisenberg models in any dimensions with $\JAF$ being replaced by $c\JAF$, where $c$ is the coordination number of a decorating $b$-spin, e.g., $c=2$ for the bond decoration~\cite{Yin_Ising_III_PRL}.  
%and will be referred to as a ``Coronaspin'' model from now on.

Regarding \textbf{the order of limits with $J\to \infty$}, one should take the thermodynamic limit
$N\to\infty$ with 
$h\ne0$ and evaluate the partition function, the free energy, and thermodynamic properties; % by the Laplace method; 
only then do we send $h\to 0^\pm$
to diagnose symmetry breaking and phase boundaries.
 
The classical macrospin $\mathbb{S}_a$ is represented by the spherical coordinates $(NS_a, \theta_a, \phi_a)$. The bath consisting of quantum $b$-spins will experience an effective field of strength
\begin{equation}
h_b(x)=\sqrt{(h\mu_b)^2+(\JAF S_a)^2-2(h\mu_b)(\JAF S_a) x},    \label{eq:hb}
\end{equation}
where $x=\cos\theta_a$. The polar angle of the $b$-spins' quantization axis satisfies $h_b \sin \theta_b=\JAF S_a \sin \theta_a$, or equivalently,
\begin{equation}
    \cos \theta_b=\frac{h\mu_b-\JAF S_a x}{h_b(x)}, \label{eq:theta_b}
\end{equation}
Their contribution to the partition function is $\left[Z_b(x)\right]^N$, where
\begin{equation}
    Z_b(x)=\sum_{m=-S_b}^{S_b}e^{\beta h_b(x)m} = \frac{\sinh[\beta(S_b+\frac{1}{2})h_b(x)]}{\sinh[\tfrac{1}{2}\beta h_b(x)]}.\label{eq:Zb}
\end{equation}
The resulting partition function is 
\begin{eqnarray}
  Z%&=&
  %2\pi\int_0^\pi e^{\beta h\mu_aNS_a\cos\theta} \left[Z_b(x)\right]^N\sin\theta d\theta \nonumber\\
  &=&
  2\pi\int_{-1}^1 \left[e^{\beta h\mu_a S_a x} Z_b(x)\right]^N dx. \label{eq:Z}   
\end{eqnarray}
%It seems that for $h=0$, $Z=4\pi\left[\frac{\sinh[\beta(S_b+\frac{1}{2})\JAF S_a]}{\sinh[\tfrac{1}{2}\beta \JAF S_a]}\right]^N$, which would imply no spontaneous transition for any dimension, as originally argued by Ising~\cite{Ising1925}. This dilemma is resolved by understanding the thermodynamic limit in such a way that the $h\to0$ limit should be taken after $N\to\infty$. Then, for 
With $h\ne 0$, the free energy per unit cell $ f=\lim\limits_{N\to\infty}-\frac{1}{N\beta}\ln Z$ is obtained by using the Laplace method~\cite{BleisteinHandelsman1986}:
\begin{eqnarray}
  f&=& \min \left\{-h\mu_a S_a x - \frac{1}{\beta}\ln Z_b(x), x\in [-1,1]\right\}. \label{eq:f}
\end{eqnarray}
Let $x^*$ be the solution of the above minimization problem, i.e., $x^*=1$, $-1$, or $
\left.\frac{\partial f}{\partial x}\right|_{x=x^*}=0$ which yields
\begin{equation}
  B_{S_b}(\beta h_b^*S_b)=\frac{\mu_a}{\mu_b}\frac{h_b^*}{\JAF S_b}, \label{eq:x_h}
\end{equation}
\ignore{\begin{eqnarray}
  h\mu_a -  S_bB_{S_b}(\beta h_b^*S_b)\frac{h\mu_b\JAF}{h_b^*}
  =0. \label{eq:x}
\end{eqnarray}}
where $h_b^*=h_b(x^*)$ and $B^{}_S(u) = \frac{2S+1}{2S}\coth\left(\frac{2S+1}{2S}u\right)- \frac{1}{2S}\coth\left(\frac{1}{2S}u\right)$ is the Brillouin function for paramagnetism. 
\ignore{\eqref{eq:x} is satisfied for 
\begin{equation}
  B_{S_b}(\beta h_b^*S_b)=\frac{\mu_a}{\mu_b}\frac{h_b^*}{\JAF S_b}. \label{eq:x_h}
\end{equation}
}

\emph{(i) $T=0$: }In the ground state, $B_{S_b}(\beta h_b^*S)=1$. \eqref{eq:x_h} is reduced to $
h_b^*=\frac{\mu_b}{\mu_a}\JAF S_b$ and \eqref{eq:theta_b} becomes $\mu_bS_b\sin\theta_b^*=\mu_aS_a\sin\theta_a^*$ for $-1<x^*<1$. These results exactly reproduce \eqref{eq:ma0} and \eqref{eq:mb0} obtained above in linear spin-wave theory for the original site-decorated quantum Heisenberg model. That $h_b^*$ is independent of $h$ in the canted phase provides a simple explanation of why $m_b^z$ remains nearly unchanged as $h$ increases [Fig.~\ref{Fig:angle}(a)].
  
\begin{figure}[t]
    \begin{center}
\includegraphics[width=\columnwidth,clip=true,angle=0]{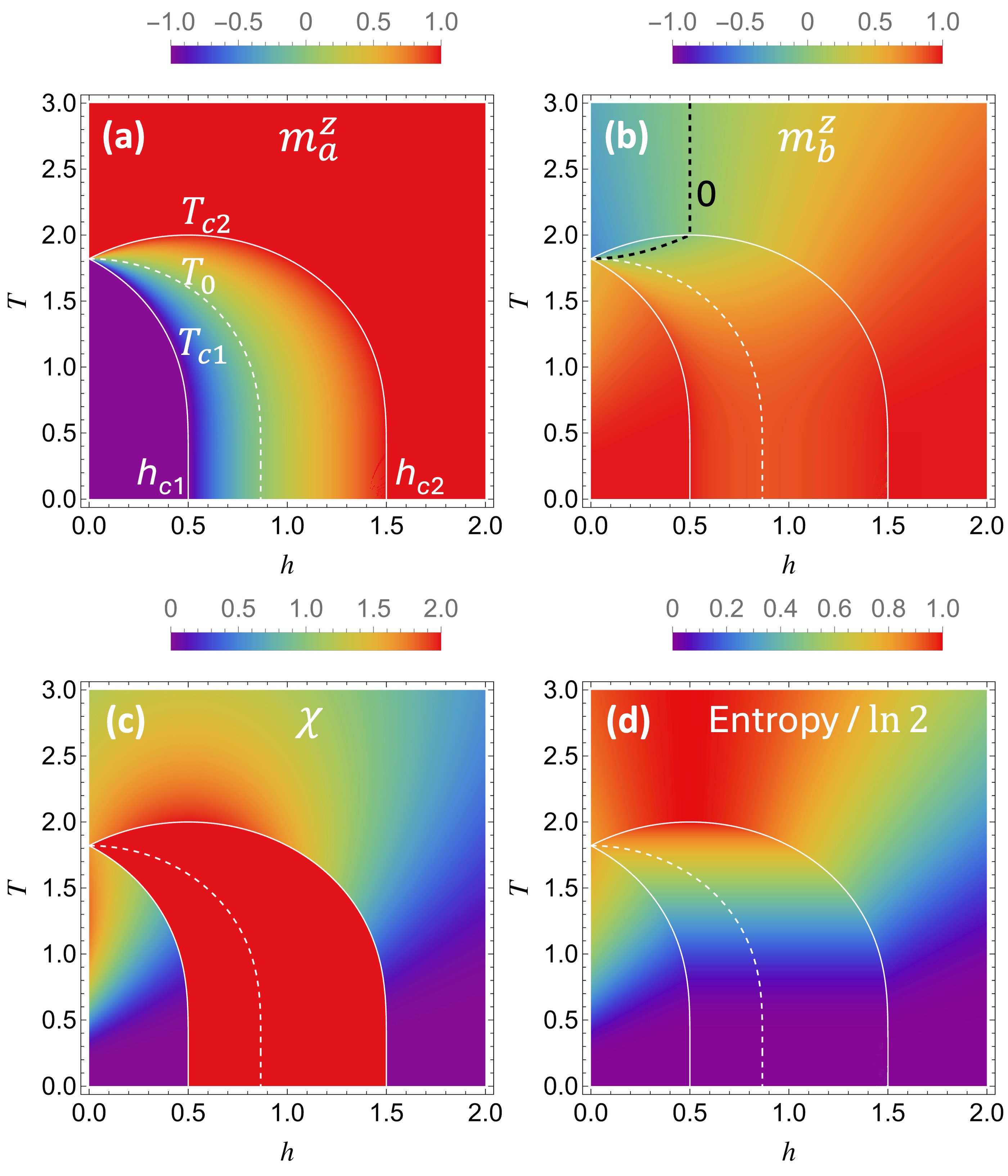}
\vspace{-0.8cm}
\ignore{
        \subfigure[][]{
\includegraphics[width=0.46\columnwidth,clip=true,angle=0]{fig3a.pdf}
        }
        \subfigure[][]{
\includegraphics[width=0.46\columnwidth,clip=true,angle=0]{fig3b.pdf}
        }
        \subfigure[][]{
\includegraphics[width=0.46\columnwidth,clip=true,angle=0]{fig3c.pdf}
        }
        \subfigure[][]{
\includegraphics[width=0.46\columnwidth,clip=true,angle=0]{fig3d.pdf}
        }
}
\ignore{
        \subfigure[][]{
\includegraphics[width=0.65\columnwidth,clip=true,angle=0]{fig3e.pdf}
        }
        \subfigure[][]{
\includegraphics[width=0.65\columnwidth,clip=true,angle=0]{fig3f.pdf}
        }
}
    \end{center}
\caption{Density plots of (a) $m_a^z$, (b) $m_b^z$, (c) $\chi$, and (d) entropy / $\ln(2S+1)$ %, (e) specific heat, (f) magnetocaloric ratio 
in the $h-T$ plane. $\newaa\equiv\JF S_a^2\to\infty$, $\newab\equiv\JAF S_aS_b=1$, $\newa\equiv\mu_aS_a=1$, $\newb\equiv\mu_bS_b=2$, $S_a=S_b=1/2$. The white solid lines indicate $T_{c1}$ and $T_{c2}$; the white dashed lines indicate $T_0$. The black dashed line in (b) is the $m_b^z=0$ contour line.}
\label{Fig:phasediagram}
\end{figure}

%\subsection{$T>0$}
\emph{(ii) Finite $T$: }There are three characteristic temperatures: $T_0$ at which $x^*=0$ and the boundary temperatures $T_{c1}$ below which $x^*=-1$ and $T_{c2}$ above which $x^*=+1$ (see Supplemental Material~\cite{SI}). The $h$ dependence of $T_{c1}$, $T_{c2}$, and $T_0$ is shown as the white lines in the finite-temperature phase diagram (Fig.~\ref{Fig:phasediagram}). They develop from the zero-temperature critical points at $h_{c1}$ and $h_{c2}$ and the $a$-spin ``flipping'' point at $\sqrt{h_{c1}h_{c2}}$, respectively. As $h\to 0$, they approach to the same temperature determined by 
\begin{eqnarray}
  B_{S_b}\left(\frac{\JAF S_aS_b}{T_{0}}\right)&=&
    \frac{h_{c2}-h_{c1}}{h_{c2}+h_{c1}}=\frac{\mu_aS_a}{\mu_bS_b}. \label{eq:T0h0}
\end{eqnarray} 

The normalized sublattice magnetizations are
\begin{subequations}
\begin{eqnarray}
  m_a^z&=&\frac{1}{NS_a}\langle \mathbb{S}^z_a\rangle
  %=-\frac{\partial f}{S_ah\partial \mu_a}
  = x^*=\cos\theta_a^*,\label{eq:ma} \\
  m_b^z&=&\frac{1}{S_b}\langle {S}^z_{i,b}\rangle=B_{S_b}(\beta h_b^*S_b)\cos\theta_b^*\label{eq:mb}
\end{eqnarray}
\end{subequations}
for the $a$- and $b$-spins, respectively, as shown in Figs.~\ref{Fig:phasediagram}(a) and \ref{Fig:phasediagram}(b). The system's total magnetization is $N m=NS_a\mu_a m_a^z+NS_b\mu_b m_b^z$ and magnetic susceptibility per $b$-spin is $\chi=\frac{\partial m}{\partial h}$ (see Supplemental Material~\cite{SI}). In the canted phase, $\chi= \mu_a\mu_b/\JAF$ is a constant, which discontinuously jumps from $\chi$ in the other two phases, as shown in Fig.~\ref{Fig:phasediagram}(c). Thus, the finite-temperature phase transitions at $T_{c1}$ and $T_{c2}$ are second-order. In comparison, for the site-decorated Ising model in the $J\to\infty$ limit, spin canting is impossible and $m_a^z=\mathrm{sgn}(T-T_0)$, resulting in a first-order transition at $T_0$.

The entropy per $b$-spin $S=-\frac{\partial f}{\partial T}$ (see Supplemental Material~\cite{SI}) normalized by its largest possible value $\ln(2S+1)$ is shown in Fig.~\ref{Fig:phasediagram}(d). The entropy is continuous, but the specific heat $C_v=T\frac{\partial S}{\partial T}$, as shown in Supplemental Fig.~S1, %Fig.~\ref{Fig:phasediagram}(e), 
is discontinuous at $T_{c1}$ and $T_{c2}$, confirming that the two transitions are second-order.

Direct comparison of the results obtained
%from the quantum limit $S_b=1/2$, where $B_{\frac{1}{2}}(x)=\tanh(x)$ and $\tanh^{-1}(x)=\frac{1}{2}\ln\left(\frac{1+x}{1-x}\right)$, to the classical spin limit $S_b\to\infty$, where the Brillouin function is replace by the Langevin function
%\begin{equation}
% $   B_{\infty}(x)\to L(x)=\coth{x}-\frac{1}{x}$,
%\end{equation} 
for different quantum spin values can be made by keeping 
\begin{eqnarray}
    \newa=\mu_aS_a,\; %\nonumber\\
    \newb=\mu_bS_b,\; %\nonumber\\
    \newab=\JAF S_aS_b
\end{eqnarray} 
fixed~\cite{Fisher_64_1D_Heisenberg_classical}. Then we found that the results are qualitatively similar; in particular, $h_{c1}$ and $h_{c2}$ remain unchanged. The main difference is that, as $S_b$ increases from $1/2$ to $\infty$, the magnitudes of $T_{c1}$, $T_{c2}$, and $T_{0}$ scale down by a factor of up to $\sim3$ for $\newa/\newb\le0.5$ because $B_S(x)\approx \frac{(S+1)}{3S} x$ for small $x$. So, it suffices to present the results for $S_b=1/2$ only.

\emph{Half-ice, half-fire state.---}This state is characterized by the $b$-spins being fully disordered with entropy $\ln(2S+1)$---``on fire''---which can be seen in Fig.~\ref{Fig:phasediagram}(d) as the red zone centered at $h=\newab/\newb$ above $T_{c2}$, where $x^*=1$---the $a$-spins are fully ordered ``ice''---and $h_b=(h\newb-\newab)/S_b=0$. That is, the $b$-spins experience a zero effective field, also resulting in $m_b^z=0$ [black dashed line in Fig.~\ref{Fig:phasediagram}(b)]. This half-ice, half-fire state is clearly invisible in the ground-state phase diagram.

For the 1D and 2D site-decorated Heisenberg models with finite $J$, the second-order transitions at $T_{c1}$ and $T_{c2}$ disappear and are replaced by one crossover from $m_a^z=-1$ to $m_a^z=+1$. The $0 < h < h_{c1}$ regime has a well-defined upper bound of the crossover width $2\delta T \le T_{c2} - T_{c1}$, which approaches zero as $h\to 0$. On the other hand, the half-ice, half-fire zone also extends well into the weak field limit. 
These results suggest that UNPC exists in low-dimensional decorated Heisenberg models in a weak field, which will be further studied below. 

%\section{$h\to 0$}
\emph{Weak field $h\mu_b\ll \JAF S_a$.---}%The conclusion that UNPC exists in weak fields in the central-macrospin model is based on the condition $J\to\infty$ followed by $h\to0$. 
Now we examine the case of weak $h$ and finite $J$ using the site-decorated Heisenberg model, \eqref{eq:model}, not the central-macrospin model. We consider the condition $JS_a^2\gg T_0$, where the $a$-spins are still locked together and treated as classical spins.

In a weak external field $h\mu_b \ll \JAF S_a$, the strength of the effective field experienced by a quantum $b$-spin $\mathbf{S}_{i,b}$, \eqref{eq:hb}, is linear in $x_i=\cos\theta_{i,a}$, where $\theta_a$ is the polar angle of the $i$th classical $a$-spin $\mathbf{S}_{i,a}$. That is, 
%\begin{eqnarray}  
$h_b(x_i)\simeq\JAF S_a -h\mu_b\,x_i$. %\label{eq:hbh0}
%\end{eqnarray}
This $b$-spin's contribution to the partition function is
\begin{eqnarray}  
Z_b(x_i)&=&\frac{\sinh[\beta(S_b+\frac{1}{2})\JAF S_a]}{\sinh[\tfrac{1}{2}\beta \JAF S_a]}\nonumber\\
&&\times\; e^{-\beta S_b B_{S_b}(\beta \JAF S_a S_b)h\mu_b x_i}.
\label{eq:Zbh0}
\nonumber
\end{eqnarray}
Therefore, the site-decorating $b$-spins can be easily summed out for any dimensions~\cite{Yin_site_arXiv}, resulting in the partition function 
\begin{equation}
  Z=\left[\frac{\sinh[\beta(S_b+\frac{1}{2})\JAF S_a]}{\sinh[\tfrac{1}{2}\beta \JAF S_a]}\right]^N\mathrm{Tr}\,e^{-\beta H_\mathrm{eff}},\label{Z_h0}    
\end{equation}
where $H_\mathrm{eff}$ is the effective Hamiltonian for the $a$-spins:
\begin{equation}
H_\mathrm{eff}=-\newaa\sum_{\langle ij\rangle}\mathbf{s}_{i,a}\cdot\mathbf{s}_{j,a}-h_\mathrm{eff}\newa\sum_i s_{i,a}^z,\label{eq:Heff}    
\end{equation}
where $\mathbf{s}_{i,a}=\mathbf{S}_{i,a}/S_a$ and $\mathbf{s}_{i,b}=\mathbf{S}_{i,b}/S_b$ are unit vectors and $\newaa=\JF S_a^2$. $H_\mathrm{eff}$ is of the same form as the undecorated (classical) Heisenberg model defined in Eq.~(\ref{eq:Ha})---with $h$ being replaced by a temperature-dependent effective magnetic field 
\begin{equation}
    h_\mathrm{eff}= h\left[1-\frac{\newb}{\newa}B_{S_b}(\beta \newab)\right]. \label{eq:heff}
\end{equation}
$h_\mathrm{eff}$ is independent of $J$. 
A phase crossover occurs at $T_0$ determined by $h_\mathrm{eff}=0$, i.e., 
%\begin{equation}
 $   B_{S_b}\left(\frac{\newab}{T_0}\right)=\frac{\newa}{\newb}$,  %\label{eq:T0h0_classical}
%\end{equation}
in agreement with \eqref{eq:T0h0} for $J\to\infty$. This must be the case, as the $J$ independence of $h_\mathrm{eff}$ means that it works for $J\to\infty$. 

\begin{figure}[t]
    \begin{center}
%        \subfigure[][]{
\includegraphics[width=\columnwidth,clip=true,angle=0]{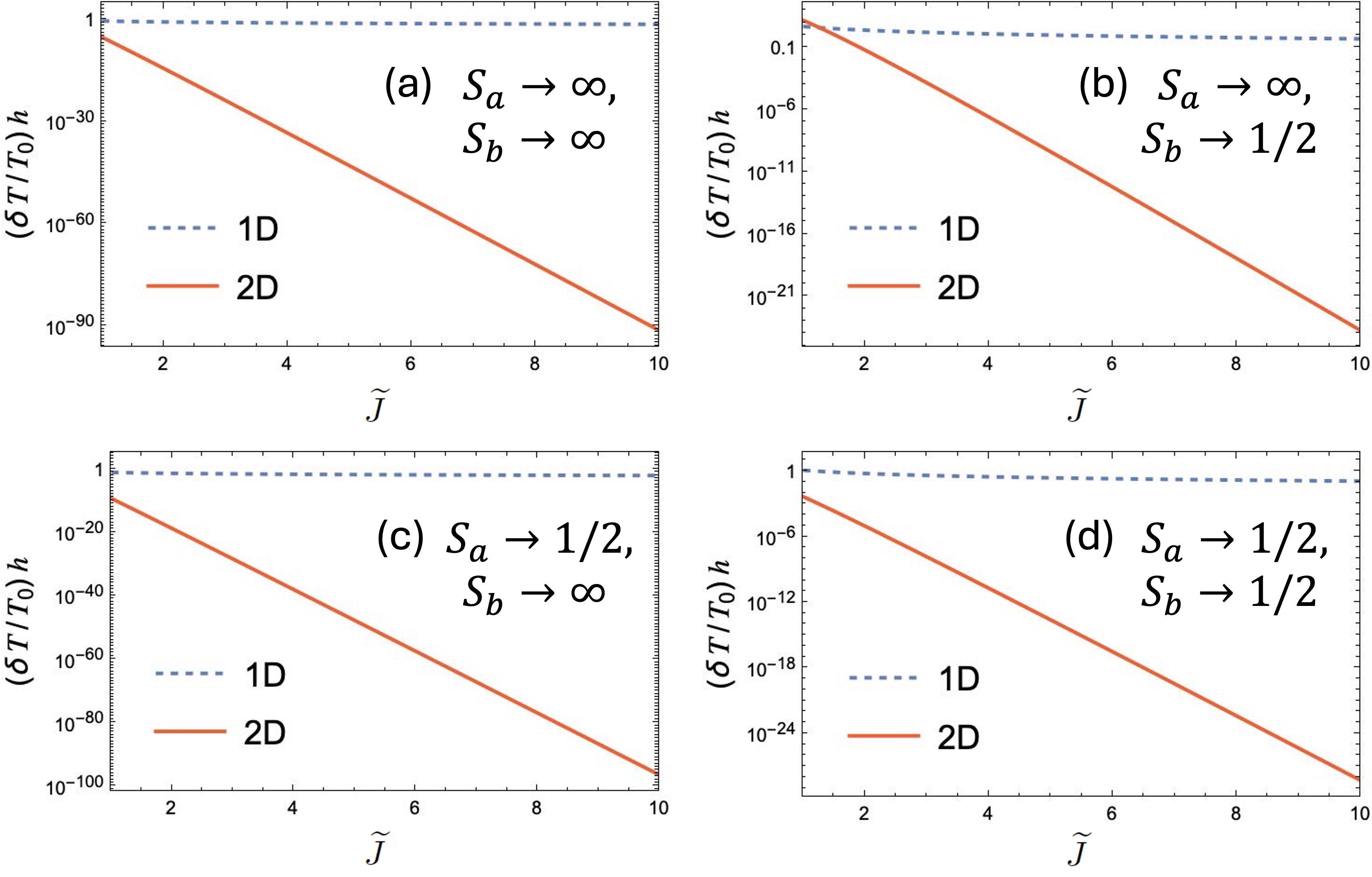}
%        }
\ignore{
        \subfigure[][]{
\includegraphics[width=0.65\columnwidth,clip=true,angle=0]{figS2.jpg}
        }
}
    \end{center}
\vspace{-0.3cm}
\caption{Estimated $\newaa$ dependence of $\frac{\delta T}{T_0} h$ of the 1D and 2D site-decorated Heisenberg models in a weak external magnetic field $h$ for (a) $S_a\to\infty$ and $S_b\to\infty$, (b) $S_a\to\infty$ and $S_b=1/2$, (c) $S_a=1/2$ and $S_b\to\infty$, and (d) $S_a=S_b=1/2$. Here $\newab=1$, $\newa=1$, and $\newb=2$, yielding $T_0\simeq 1.82$ for $S_b=\frac{1}{2}$ and $T_0\simeq 0.56$ for $S_b\to\infty$.}
\label{Fig:chi}
\end{figure}

%Although Eq.~(\ref{eq:Heff}) has not been solved explicitly in the presence of a finite magnetic field, for which the 1D classical ferromagnetic chain has been numerically studied~\cite{Blume_PRB_75_1D_Classical_Heisenberg_field}, we can still retrieve accurate information about 
The crossover width $2\delta T$ can be estimated via 
%\begin{eqnarray}
    $\delta T=\left(\frac{\partial \langle s^z_{i,a}\rangle}{\partial T}\right)^{-1}_{T=T_0}=\left(\frac{\partial \langle s^z_{i,a}\rangle}{\partial h_\mathrm{eff}}\right)^{-1}_{h_\mathrm{eff}=0}\left(\frac{\partial h_\mathrm{eff}}{\partial T}\right)^{-1}_{T=T_0}$~\cite{Yin_site_arXiv,Yin_MPT_chain,Yin_MPT,Yin_Ising_III_PRL}:
%\end{eqnarray}
%One finds
\begin{eqnarray}
    \delta T
    %&=&\left(\frac{\chi_0}{\newa}\right)^{-1}\left(\frac{\partial h_\mathrm{eff}}{\partial T}\right)^{-1}_{T=T_0} \nonumber\\
    &=&\left[\frac{\chi_0h\newb}{T_0^2\newa^2}\newab B'_{S_b}\left(\frac{\newab}{T_0}\right)\right]^{-1},
    \label{eq:dT}
\end{eqnarray}
where $\chi_0=\newa\left.\frac{\partial \langle s^z_{i,a}\rangle}{\partial h_\mathrm{eff}}\right|_{h_\mathrm{eff}=0}$ is the initial magnetic susceptibility of the effective Hamiltonian, Eq.~(\ref{eq:Heff}), at $T_0$. The known accurate results of $\chi_0$ for different dimensions and different $S_a$, $S_b$ values are listed in Table~\ref{table:chi}~\cite{SI}. 
\ignore{
\begin{equation}
    \chi_0=\left\{
\begin{array}{ll}  
    \frac{2\newa^2\newaa}{3T_0^2}& \mathrm{for\; 1D\; chain},\\
    \frac{1}{128\pi e^\pi}\frac{T_0}{3\newaa^2} e^{\frac{4\pi\newaa}{T_0}}& \mathrm{for\; square\; lattice}.\\
\end{array}
\right.
\label{eq:chi0}
\end{equation}

$\chi_0=\frac{2\newa^2\newaa}{3T_0^2}$ for a 1D chain~\cite{Fisher_64_1D_Heisenberg_classical} and $\chi_0=\frac{\newa^2}{128\pi e^\pi}\frac{T_0}{3\newaa^2}    e^{\frac{4\pi\newaa}{T_0}}$ for a square lattice~\cite{Takahashi_PRB_87_2D-Classical-Heisenberg}.} Therefore, 
\begin{eqnarray}
    \frac{\delta T}{T_0}h \propto \left\{
\begin{array}{ll}  
    \frac{T_0}{\newaa}& \mathrm{for\; 1D\; chain},\\
    e^{-\frac{4\pi\newaa}{T_0}}& \mathrm{for\; square\; lattice}.\\
\end{array}
\right.
\label{eq:dT}
\end{eqnarray}
Increasing $\newaa$ does not affect $T_0$ but narrows $2\delta T$. As shown in Fig.~\ref{Fig:chi}, the power-law decay of $\frac{\delta T}{T_0}$ as a function of $\newaa$ makes the narrowing inefficient in the 1D chain. By sharp contrast, even for modest $\newaa\sim5$, the exponential decay of $\frac{\delta T}{T_0}$ to an extremely small value in the square lattice signals UNPC in two dimensions. %with classical $a$-spins and quantum $b$-spins, e.g., $\frac{\delta T}{T_0}\simeq \frac{10^{-9}}{h}$ for a modest $\newaa=5$ with $\newab=1$, $\newa=1$, and $\newb=2$ yielding $T_0=1.82$. In comparison, $\frac{\delta T}{T_0}\simeq \frac{10^{-43}}{h}$ when both $a$- and $b$-spins are classical due to a smaller $T_0=0.56$. 
This echoes the emergence of UNPC in 1D decorated Ising ferrimagnets, where $\chi_0\propto \frac{1}{T_0}e^{\frac{2J}{T_0}}$, resulting in $\frac{\delta T}{T_0}\propto e^{-\frac{2J}{T_0}}$~\cite{Yin_Ising_III_PRL} or $\frac{\delta T}{T_0}h\propto e^{-\frac{2J}{T_0}}$ in a weak field~\cite{Yin_site_arXiv}. 

We emphasize that the present results are exact only in the $J\to\infty$
CMM; finite-$J$ results are predictions from a rigorous weak-field mapping for classical 
$a$-spins plus known $\chi_0$. Since quantum and classical $\chi_0$'s share the same 
$e^{4\pi\newaa/T_0}$ factor in 2D, we expect the same exponential narrowing when both spins are quantum. %(Table~\ref{table:chi})~\cite{Takahashi_PRL_87_1D2D-quantum-Heisenberg,Takahashi_PTPS_86_1D2D_quantum_Heisenberg,Takahashi_PRB_87_2D-Classical-Heisenberg}, the present results strongly suggest that UNPC exist in decorated 2D Heisenberg ferrimagnets where both $a$- and $b$-spins are quantum for $e^{-4\pi\newaa/T_0}\ll 1$. 
Verifications by future massive-scale computer simulations are highly desirable.   

%Finally, for higher-dimensional site-decorated Heisenberg models, where a long-range spin order exists at $T_c>T_0$, the macrospin model is still valid for $J\to\infty$, which means $T_c\to\infty$, in agreement with the result that $m_a^z=1$ for $T>T_{c2}$, unlike the Ising case where an $a$-spin reversal transition occurs at $T_0(h)$~\cite{Yin_site_arXiv}.

In summary, we have presented a study of the site-decorated quantum Heisenberg model in two limits: (i) the large $J$ limit, where the system is described by the exactly solvable central-macrospin model, featuring two second-order transitions at $T_{c1}$ and $T_{c2}$, which meet at $T_0$ in the weak field limit, to a half-ice, half-fire state; (ii) the weak field limit with finite $J$, where the system is mapped onto a simple Heisenberg model in an effective field, which changes sign at $T_0$. We predict that an ultranarrow phase crossover, driven by the half-ice, half-fire state, exists at $T_0$ %$T_0=\JAF S_aS_b/B_{S_b}^{-1}\left(\frac{\mu_aS_a}{\mu_bS_b}\right)$ 
in 2D decorated Heisenberg ferrimagnets in a weak external magnetic field thanks to its exponentially strong magnetic susceptibility $\chi_0\propto e^{4\pi JS_a^2/T_0}$, though less likely in 1D chains where $\chi_0\propto J/T_0$. We anticipate these results to shed light on new technological applications of low-dimensional quantum spin systems and attract experimental and computational tests. Decorated optical lattices~\cite{Bernien_Nature_17_Rydberg} and  $d$-$f$ compounds~\cite{Ramirez_25_SmMn2Ge2} are possible real systems to demonstrate the half-ice, half-fire driven UNPC phenomenon. 

%\section*{Acknowledgments}
The work at Brookhaven National Laboratory was supported by the U.S. Department of Energy (DOE), Division of Materials Science, under Contract No. DE-AC02-98CH10886. 

%\bibliography{Ising_quantum,spin-bath}

\beginsupplement
\onecolumngrid
\newpage
%\beginsupplement
%\onecolumngrid
%\newpage
%\input{SI}
\begin{center}
\textbf{\large [Supplemental Material] \mytitle\\}

\textrm{\\Weiguo Yin and A. M. Tsvelik\\}
%\email{wyin@bnl.gov}
\vspace{0.1cm}
\textsl{Condensed Matter Physics and Materials Science Division, Brookhaven National Laboratory, Upton, New York 11973, USA\\}
%\textrm{(Received 31 March 2025; revised 6 June 2025; accepted 26 August 2025; published 11 September 2025)}
\end{center}
\vspace{1cm}

\twocolumngrid

\section{Ground-state phase diagram}

\begin{figure}[h]
 \begin{center}
\includegraphics[width=\columnwidth,clip=true,angle=0]{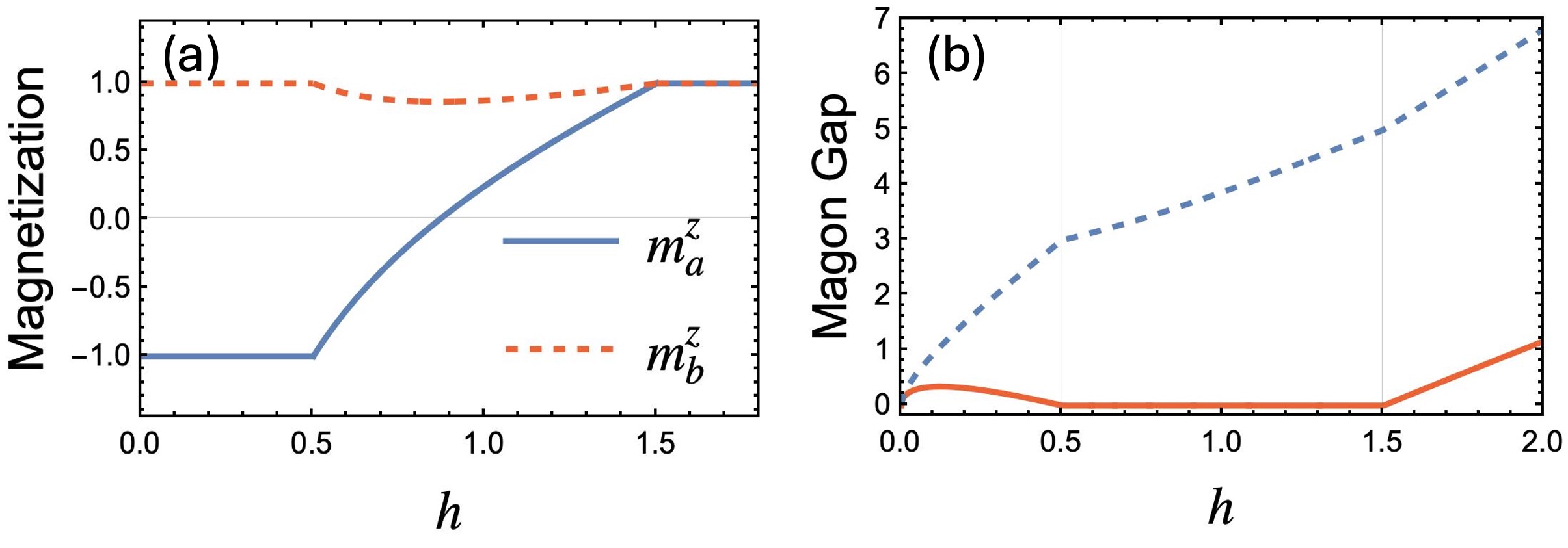}
\vspace{-0.8cm}
 \ignore{   
        \subfigure[][]{
\includegraphics[width=0.49\columnwidth,clip=true,angle=0]{Fig2a.jpeg}
        }
}
    \end{center}
\caption{Zero-temperature magnetic field dependence of (a) $m_a^z=\cos\theta_a^*$ (blue solid line) and $m_b^*=\cos\theta_b^*$ (red dashed line), and of (b) the spectral gaps of the acoustic (red solid line) and optic branches (blue dashed line) in linear spin-wave theory. $z\newaa\equiv z\JF S_a^2=10$ with $z$ being the coordination number of the $a$-spins within the backbone, $\newab\equiv\JAF S_aS_b=1$, $\newa\equiv\mu_aS_a=1$, $\newb\equiv\mu_bS_b=2$, $S_a=S_b=1/2$. Vertical grid lines in (b) mark $h_{c1}=0.5$ and $h_{c2}=1.5$.}
\label{Fig:angle}
\end{figure}

The zero-temperature phase diagram of the site-decorated Heisenberg model described by Eqs.~(\ref{eq:ma0})--(\ref{eq:mb0}) is plotted in Fig.~\ref{Fig:angle}(a). The spectral gaps in linear spin-wave theory is plotted in Fig.~\ref{Fig:angle}(b).  

\section{Analytical derivations}

There are three characteristic temperatures: $T_0$ at which $x^*=0$ and the boundary temperatures $T_{c1}$ below which $x^*=-1$ and $T_{c2}$ above which $x^*=+1$. They are determined by 
\begin{subequations}
\begin{eqnarray}
  B_{S_b}\left(\frac{h_b^-S_b}{T_{c1}}\right) &=& \frac{2h-h_{c1}+h_{c2}}{h_{c1}+h_{c2}},\label{eq:Tc1}\\
  B_{S_b}\left(\frac{h_b^+S_b}{T_{c2}}\right) &=& \frac{2h+h_{c1}-h_{c2}}{h_{c1}+h_{c2}},
  \label{eq:Tc2}\\
  B_{S_b}\left(\frac{h_b^0S_b}{T_{0}}\right)&=&
    %\frac{\mu_a}{\mu_b}\frac{h_b^0}{\JAF S_b}=
    \sqrt{1+\frac{4(h^2-h_{c1}h_{c2})}{(h_{c1}+h_{c2})^2}}, \label{eq:T0}
\end{eqnarray}
\end{subequations}
where $h_b^\mp=h\mu_b\pm \JAF S_a$ and $h_b^0=\sqrt{(h\mu_b)^2+(\JAF S_a)^2}$. Since $|B_{S_b}(x)|< 1$, finite $T_{c1}$, $T_{c2}$, and $T_0$ exist for $h<h_{c1}$, $h<h_{c2}$, and $h<\sqrt{h_{c1}h_{c2}}$ respectively. The $h$ dependence of $T_{c1}$, $T_{c2}$, and $T_0$ is shown as the white lines in the finite-temperature phase diagram (Fig.~\ref{Fig:phasediagram}).

Therefore, a quick picture of regimes emerges: 
%(AFM $J=-1$, $\mu_a=1, \mu_b=2; h>0)
% Temperatures are in units of |J|/k_B.
\begin{itemize}
  \item \textbf{$0 < h < h_{c1}$:}
  \begin{itemize}
    \item Low $T$: $x^*=-1$.
    \item At $T_{c1}(h)$: minimizer, \eqref{eq:f}, enters the interior.
    \item At $T_0(h)$: $x^*=0$ (so $m^z_a=0$).
    \item At $T_{c2}(h)$: minimizer is captured by $x^*=+1$.
  \end{itemize}

  \item \textbf{$h_{c1} \le h < \sqrt{h_{c1}h_{c2}}$:}
  \begin{itemize}
    \item No $T_{c1}$.
    \item Low $T$: interior minimizer.
    \item At $T_0(h)$: $x^*=0$.
    \item At $T_{c2}(h)$: captured by $x^*=+1$.
  \end{itemize}

  \item \textbf{$\sqrt{h_{c1}h_{c2}} \le h < h_{c2}$:}
  \begin{itemize}
    \item Low $T$: interior minimizer with $x^*>0$.
    \item No $T_0$.
    \item At $T_{c2}(h)$: captured by $x^*=+1$.
  \end{itemize}

  \item \textbf{$h \ge h_{c2}$:}
  \begin{itemize}
    \item $x^*(T,h)\equiv 1$ for all $T$ (no $T_{c2}$, no $T_0$).
  \end{itemize}
\end{itemize}

The system's total magnetization is $N m=NS_a\mu_a m_a^z+NS_b\mu_b m_b^z$ and magnetic susceptibility per $b$-spin $\chi=\frac{\partial m}{\partial h}$ is given by
\begin{eqnarray}
    \chi=\left\{
\begin{array}{cl}
    \beta(\mu_bS_b)^2B_{s_b}'(\beta h_b^\pm S_b) & \mathrm{for}\;\; x=\mp1,\\
     \mu_a\mu_b/\JAF & \mathrm{for}\;\; x\in(-1,1),\\
\end{array}\right.\label{eq:chi} 
\end{eqnarray}
\vspace{0.1cm}where $B_S'(u)=\frac{\partial B_S(u)}{\partial x}=-\left(\frac{2 S + 1}{2 S}\right)^2 \csch^2\left(\frac{2 S + 1}{2 S}u\right) +\left(\frac{1}{2 S}\right)^2 \csch^2\left(\frac{1}{2 S}u\right)$. 
We have used \eqref{eq:theta_b} and \eqref{eq:x_h} in deriving \eqref{eq:chi}. In the canted phase, $\chi$ is a constant, which discontinuously jumps from $\chi$ in the other two phases, as shown in Fig.~\ref{Fig:phasediagram}(c). Thus, the finite-temperature phase transitions at $T_{c1}$ and $T_{c2}$ are second-order.

The entropy per $b$-spin $S=-\frac{\partial f}{\partial T}$ is given by
\begin{equation}
    S=\ln{Z_b(x^*)}-\frac{h_b^*S_b}{T}B_{S_b}(\beta h_b^*S_b),
\end{equation}
as shown in Fig.~\ref{Fig:phasediagram}(d). 

\begin{figure}[h]
    \begin{center}
%        \subfigure[][]{
\includegraphics[width=0.65\columnwidth,clip=true,angle=0]{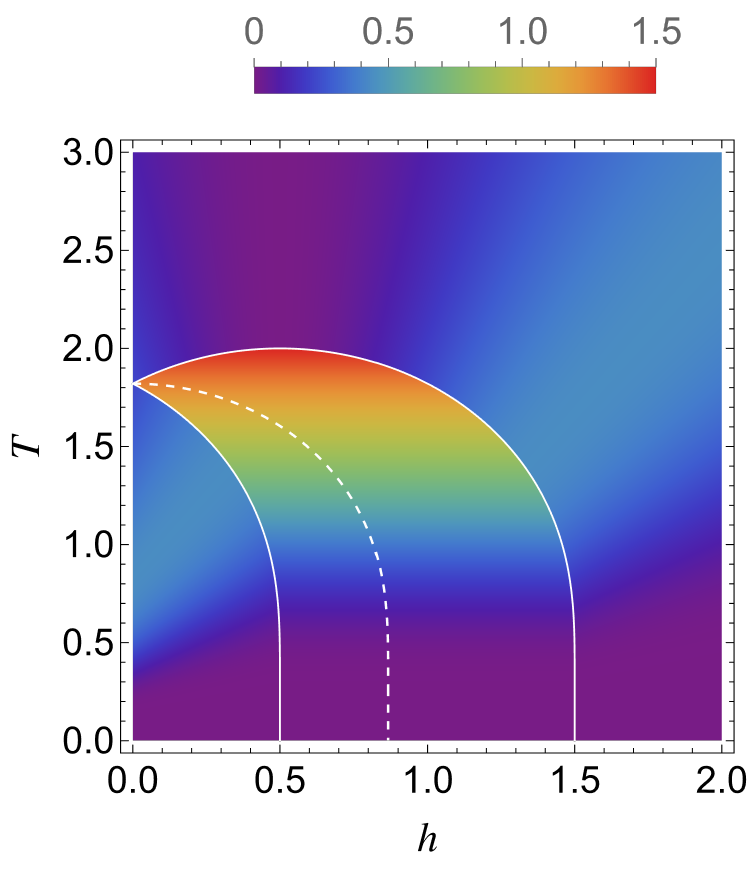}
%        }
\ignore{
        \subfigure[][]{
\includegraphics[width=0.65\columnwidth,clip=true,angle=0]{fig3f.pdf}
        }
}
    \end{center}
\vspace{-0.3cm}
\caption{Density plots of specific heat
in the $h-T$ plane. $\JF\to\infty$, $\newab\equiv\JAF S_aS_b=1$, $\newa\equiv\mu_aS_a=1$, $\newb\equiv\mu_bS_b=2$, $S_a=S_b=1/2$. The white solid lines indicate $T_{c1}$ and $T_{c2}$; the white dashed lines indicate $T_0$.}
\label{Fig:Cv}
\end{figure}

The entropy is continuous, but, as shown in %Supplemental 
Fig.~\ref{Fig:Cv}, %Fig.~\ref{Fig:phasediagram}(e), 
the specific heat $C_v=T\frac{\partial S}{\partial T}$ given by
\begin{eqnarray}
    C_v=\left\{
\begin{array}{cl}
    (\beta h_b^\pm S_b)^2B_{s_b}'(\beta h_b^\pm S_b) & \mathrm{for}\;\; x=\mp1,\\
     \frac{y^2 B'_{S_b}(y) B_{S_b}(y)}{B_{S_b}(y) -y B'_{S_b}(y)} & \mathrm{for}\;\; x\in(-1,1),\\
\end{array}\right.\label{eq:Cv} \nonumber
\end{eqnarray}
where $y=\beta h_b^* S_b$, is discontinuous at $T_{c1}$ and $T_{c2}$, confirming that the two transitions are second-order.

Closed-form solutions are achieved for the quantum limit $S_b=1/2$, where $B_{\frac{1}{2}}(u)=\tanh(u)$, $B'_{\frac{1}{2}}(u)=\mathrm{sech}^2(u)$, $B^{-1}_{\frac{1}{2}}(u)=%\tanh^{-1}(u)=
\frac{1}{2}\ln\left(\frac{1+u}{1-u}\right)$, and $Z_b(x)=2\cosh[\beta h_b(x)]$. %Note that, for the classical spin limit $S_b\to\infty$, the Brillouin function is replace by the Langevin function $B_{\infty}(x)\to L(x)=\coth{x}-\frac{1}{x}$.

\begin{table}[t]
\begin{center}
\caption{The zero-field magnetic susceptibility $\chi_0$ of the standard ferromagnetic Heisenberg model for the backbone $a$-spins in one and two dimensions. $T_0$ given by \eqref{eq:T0h0} is independent of spatial dimension and determined by the decorating $b$-spins, e.g., $\newab=1$, $\newa=1$, and $\newb=2$ yield $T_0\simeq 1.82$ for $S_b=\frac{1}{2}$ and $T_0\simeq 0.56$ for $S_b\to\infty$. We also showcase the values of $\chi_0$ and $\frac{\delta T}{T_0} h$ for these parameters together with $\newaa=5$, where the numbers inside the parentheses are the values of $S_b$.}
\begin{tabular}{|c|c|c|}
\hline\hline
$S_a$ & 1D & 2D \\
\hline
& & \\
$S_a\to \infty$  &$\chi_0=\frac{2\newa^2\newaa}{3T_0^2}$~\cite{Fisher_64_1D_Heisenberg_classical} & $\chi_0=\frac{\newa^2}{128\pi e^\pi}\frac{T_0}{3\newaa^2}e^{\frac{4\pi\newaa}{T_0}}$~\cite{Takahashi_PRB_87_2D-Classical-Heisenberg}\\
& & \\
& 10.8($\infty$)%\footnote{}
, 1.0($\frac{1}{2}$) & $8.5\times10^{42}$($\infty$), $2.5\times10^9$($\frac{1}{2}$)\\
& & \\
$\frac{\delta T}{T_0} h$& 0.1($\infty$)%\footnote{}
, 1.2($\frac{1}{2}$) & $1.7\times10^{-43}$($\infty$), $4.8\times10^{-10}$($\frac{1}{2}$)\\
& & \\
\hline
& & \\
$S_a=\frac{1}{2}$ & $\chi_0=\frac{8\newa^2\newaa}{3T_0^2}$~\cite{Takahashi_PTPS_86_1D2D_quantum_Heisenberg} & $\chi_0=\frac{\newa^2}{3\pi \newaa  S_a}    e^{\frac{4\pi\newaa}{T_0}}$\hspace{0.6cm}~\cite{Takahashi_PTPS_86_1D2D_quantum_Heisenberg} \\
& & \\
& 43.0($\infty$), 4.0($\frac{1}{2}$) & $4.5\times10^{47}$($\infty$), $4.1\times10^{13}$($\frac{1}{2}$)\\
& & \\
$\frac{\delta T}{T_0} h$ & 0.03($\infty$), 0.3($\frac{1}{2}$) & $3.2\times10^{-48}$($\infty$), $2.9\times10^{-14}$($\frac{1}{2}$) \\
& & \\
\hline\hline
\end{tabular}
\label{table:chi}
\end{center}
\end{table}

\section{Estimation of The Crossover Width $2\delta T$}

The accurate information about the zero-field magnetic susceptibility $\chi_0$ of the standard ferromagnetic Heisenberg model for either quantum or classical backbone $a$-spins, \eqref{eq:Ha} and \eqref{eq:Heff}, in both the 1D chain and the 2D square lattice are known and listed in Table~\ref{table:chi}.  We also showcase the values of $\chi_0$ and $\frac{\delta T}{T_0} h$ for $\newaa=5$, $\newab=1$, $\newa=1$, and $\newb=2$. $T_0$ is independent of $\newaa$ and the spatial dimension; in the weak $h$ limit, $T_0\simeq 1.82$ for $S_b=\frac{1}{2}$ and $T_0\simeq 0.56$ for $S_b\to\infty$. 
These results strongly suggest that UNPC exist in decorated 2D Heisenberg ferrimagnets, though less likely in 1D quantum spin chains.

\section{AI and Computational tools}

%\texttt{ChatGPT 5 Thinking} was used to reshape the research, derive the equations, and translate the human-AI conversations to a Wolfram Mathematica program---the AI's contribution is ranked at 6d (AI-guided pivot) in a 9-level rating system~\cite{Yin_Potts_UNPC}. 
\texttt{ChatGPT 5 Thinking} was used to reshape the research, derive the equations, and translate the human-AI conversations into Wolfram Mathematica code. The AI's contribution is ranked on the level of AI methodology advisor (6d) according to a nine-level rating system~\cite{Yin_Potts_UNPC}. 

\texttt{ChatGPT 5} was used to generate Fig.~1(b). \texttt{Wolfram Mathematica 14.3} was used to produce Figs.~2, 3, S1, and S2. All derivations and results have been cross-verified by three methods: by hand, by the AI, and via derivatives with Mathematica, e.g., the calculation of $\chi$ via  $\chi=\frac{\partial m}{\partial h}$ vs \eqref{eq:chi}.

\end{document}